\documentclass[twocolumn]{webofc}

\usepackage[varg]{txfonts}   
\usepackage{hyperref}
\usepackage{url}
\usepackage{bm}
\hypersetup{colorlinks=true,citecolor=blue,urlcolor=blue,linkcolor=blue}
\begin{document}
\title{Constraining the nuclear equation of state from terrestrial experiments and neutron star observations using relativistic mean-field models}
%
%

\author{
  \firstname{Tsuyoshi} \lastname{Miyatsu}\inst{1}\fnsep\thanks{\email{tsuyoshi.miyatsu@ssu.ac.kr}} \and
  \firstname{Myung-Ki} \lastname{Cheoun}\inst{1}\fnsep\thanks{\email{cheoun@ssu.ac.kr}}            \and
  \firstname{Kyungsik} \lastname{Kim}\inst{2}\fnsep\thanks{\email{kyungsik@kau.ac.kr}}             \and
  \firstname{Koichi}   \lastname{Saito}\inst{3}\fnsep\thanks{\email{koichi.saito@rs.tus.ac.jp}}
}

\institute{
  Department of Physics and OMEG Institute, Soongsil University, Seoul 06978, Republic of Korea \and
  School of Liberal Arts and Sciences, Korea Aerospace University, Goyang 10540, Republic of Korea \and
  Department of Physics and Astronomy, Faculty of Science and Technology, Tokyo University of Science, Noda 278-8510, Japan
}

\abstract{We investigate the nuclear equation of state (EoS) for isospin-asymmetric matter using a new set of RMF interactions with the $\sigma$-$\delta$ and $\omega$-$\rho$ mixing, referred to as the OMEG family. These interactions are optimized so as to reproduce both terrestrial nuclear measurements and astrophysical constraints extracted from NICER and GW170817. The $\sigma$-$\delta$ mixing softens the nuclear symmetry energy and pressure around twice the saturation density, which enables relatively small neutron-star radii and tidal deformabilities while keeping the nuclear EoS sufficiently stiff at high densities to support $2M_{\odot}$ neutron stars. We find that the curvature parameter, $K_{\textrm{sym}}$, plays an important role in realizing the soft-to-hard behavior of the nuclear EoS, and the astrophysical data favor small or even negative values of $K_{\textrm{sym}}$.
}
\maketitle
\section{Introduction}
\label{sec:introduction}

Recent precise measurements of neutron-star radii by the Neutron Star Interior Composition ExploreR (NICER)~\cite{Vinciguerra:2023qxq,Choudhury:2024xbk,Salmi:2024aum,Salmi:2024bss}, together with the direct detection of gravitational-wave signals from the binary neutron-star merger event, GW170817, by Advanced LIGO and Advanced Virgo detectors~\cite{LIGOScientific:2018cki}, have significantly intensified theoretical investigations of the nuclear equation of state (EoS) for dense matter.
These multi-messenger observations are particularly valuable for clarifying the properties of isospin-asymmetric nuclear matter at high densities, which cannot be accessed through terrestrial experiments.

In this work, we introduce a new effective interaction within the relativistic mean-field (RMF) model that incorporates isoscalar- and isovector-meson mixing, namely the $\sigma^{2}\bm{\delta}^{2}$ and $\omega_{\mu}\omega^{\mu}\bm{\rho}_{\nu}\cdot\bm{\rho}^{\nu}$ terms, in order to simultaneously reproduce terrestrial nuclear data and astrophysical observations of neutron stars~\cite{Zabari:2018tjk,Miyatsu:2022wuy,Li:2022okx}.
The $\delta$ meson, an isovector-scalar meson, induces a splitting of the effective nucleon masses in asymmetric nuclear matter and influences the density dependence of the nuclear symmetry energy, $E_{\textrm{sym}}(\rho_{B})$, at high densities.

In particular, we examine whether massive neutron stars with both relatively small radii and tidal deformabilities can be explained simultaneously, while still accounting for the large neutron skin thickness of $^{208}$Pb reported by the PREX-2 experiment~\cite{PREX:2021umo}.
We further investigate how the $\delta$ meson and the $\sigma$-$\delta$ mixing influence the behavior of the EoS for isospin-asymmetric nuclear matter.

\begin{table*}
  \centering
  \caption{Coupling constants and selected properties of isospin-asymmetric nuclear matter for the OMEG family. The parameter $g_{2}$ is in units of fm$^{-1}$. The nuclear symmetry energy, $E_{\textrm{sym}}(\rho_{0})$, and its slope and curvature parameters, $L$ and $K_{\textrm{sym}}$, are given in MeV.}
  \label{tab:CCs}
  \begin{tabular}{cccccccccccccc}
    \hline
    Model & $g_{\sigma}^{2}$ & $g_{\omega}^{2}$ & $g_{\delta}^{2}$ & $g_{\rho}^{2}$ & $g_{2}$ & $g_{3}$ & $c_{3}$ & $\Lambda_{\sigma\delta}$ & $\Lambda_{\omega\rho}$ & $E_{\textrm{sym}}(\rho_{0})$ & $L$ & $K_{\textrm{sym}}$ & $\Lambda_{1.4}$ \\
    \hline
    OMEG0 & $89.4$ & $142.8$ & $37.7$ & $51.7$ &  $10.0$ &  $-21.5$ &      -- & $87.0$ &  $102.6$ & $34.6$ & $50.0$ &  $-384.4$ & $498$ \\
    OMEG1 & $99.6$ & $166.3$ & $30.0$ & $44.6$ & $~~7.8$ & $~~-1.1$ & $100.0$ & $95.0$ & $~~75.7$ & $35.1$ & $70.0$ &  $-218.8$ & $515$ \\
    OMEG2 & $99.6$ & $166.3$ & $20.0$ & $44.4$ & $~~7.8$ & $~~-1.1$ & $100.0$ & $85.0$ &  $288.9$ & $33.0$ & $45.0$ &  $-216.7$ & $458$ \\
    OMEG3 & $99.7$ & $166.3$ & $15.0$ & $57.6$ & $~~7.8$ & $~~-1.1$ & $100.0$ & $70.0$ &  $909.8$ & $30.0$ & $20.0$ & $~~-65.9$ & $462$ \\
    \hline
  \end{tabular}
\end{table*}
\begin{figure*}
\centering
\includegraphics[width=12.7cm,clip]{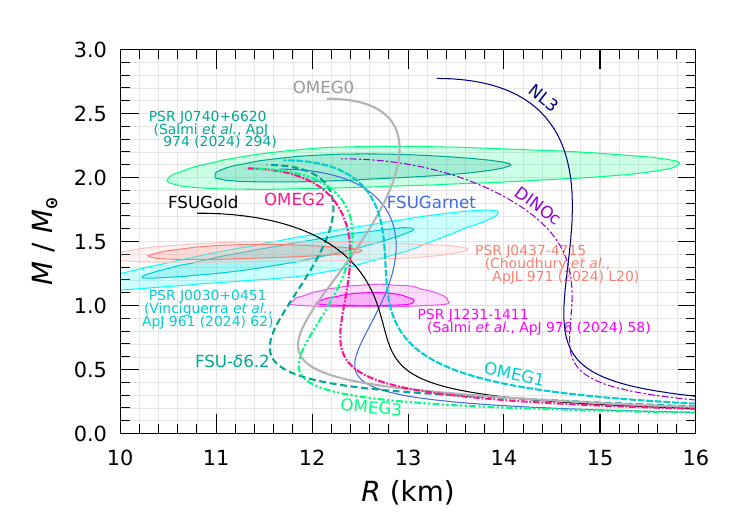}
\caption{Mass{\textendash}radius relations of neutron stars for the OMEG family. The observational constraints are taken from PSR J0030$+$0451 ($1.40^{+0.13}_{-0.12}$~$M_{\odot}$ and $11.71^{+0.88}_{-0.83}$~km)~\cite{Vinciguerra:2023qxq}, PSR J0437{\ensuremath{-}}4715 ($1.418\pm0.037$~$M_{\odot}$ and $11.36^{+0.95}_{-0.63}$~km)~\cite{Choudhury:2024xbk}, PSR J0740$+$6620 ($2.073^{+0.069}_{-0.069}$~$M_{\odot}$ and $12.49^{+1.28}_{-0.88}$~km)~\cite{Salmi:2024aum}, and PSR J1231{\ensuremath{-}}1411 ($1.04^{+0.05}_{-0.03}$~$M_{\odot}$ and $12.6\pm0.3$~km)~\cite{Salmi:2024bss}. The other theoretical results are explained in Ref.~\cite{Miyatsu:2024ioc}.}
\label{fig:1}
\end{figure*}

\section{Theoretical framework}
\label{sec:framework}

We employ an updated version of the RMF Lagrangian density, which includes the isoscalar ($\sigma$ and $\omega^{\mu}$) and isovector ($\bm{\delta}$ and $\bm{\rho}^{\mu}$) mesons, as well as nucleons ($N$)~\cite{Miyatsu:2022wuy}.
The interacting Lagrangian density is given by
\begin{align}
  \mathcal{L}_{\textrm{int}}
  & = \bar{\psi}_{N} \left[ g_{\sigma}\sigma
    - g_{\omega}\gamma_{\mu}\omega^{\mu}
    + g_{\delta}\bm{\delta}\cdot\bm{\tau}_{N}
    - g_{\rho}\gamma_{\mu}\bm{\rho}^{\mu}\cdot\bm{\tau}_{N} \right] \psi_{N} \nonumber \\
  & - U_{\textrm{NL}}(\sigma,\omega,\bm{\delta},\bm{\rho}),
\end{align}
where $\psi_{N}$ is the nucleon field and $\bm{\tau}_{N}$ is its isospin matrix.
The meson-nucleon coupling constants are respectively denoted by $g_{\sigma}$, $g_{\omega}$, $g_{\delta}$, and $g_{\rho}$.
The nonlinear potential takes the form
\begin{align}
  U_{\textrm{NL}}(\sigma,\omega,\bm{\delta},\bm{\rho})
  & = \frac{1}{3}g_{2}\sigma^{3}
    + \frac{1}{4}g_{3}\sigma^{4}
    - \frac{1}{4}c_{3}(\omega_{\mu}\omega^{\mu})^{2} \nonumber \\
  & - \Lambda_{\sigma\delta}\sigma^{2}\bm{\delta}^{2}
    - \Lambda_{\omega\rho}(\omega_{\mu}\omega^{\mu})(\bm{\rho}_{\nu}\cdot\bm{\rho}^{\nu}),
    \label{eq:NL}
\end{align}
with three coupling constants, $g_{2}$, $g_{3}$, and $c_{3}$, and two mixing parameters, $\Lambda_{\sigma\delta}$ and $\Lambda_{\omega\rho}$.
The isoscalar- and isovector-meson mixing affects only the properties of $N\not=Z$ finite nuclei and isospin-asymmetric nuclear matter~\cite{Miyatsu:2013yta,Miyatsu:2023lki}.

We present a new set of effective interactions, referred to as the OMEG family, constructed within the RMF framework with the $\sigma$-$\delta$ and $\omega$-$\rho$ mixing.
The model optimization is performed in the usual manner so as to fit the experimental data for binding energies per nucleon and charge radii of several finite, closed-shell nuclei, while properly taking into account the Coulomb interaction.
The isoscalar mixing parameter, $\Lambda_{\sigma\delta}$, is additionally determined to satisfy the astrophysical constraints on the radius and dimensionless tidal deformability of a canonical neutron star, $R_{1.4}$ and $\Lambda_{1.4}$, as inferred from NICER and GW170817~\cite{Miyatsu:2023lki,Miyatsu:2024ioc}.
The resulting coupling constants and selected properties of isospin-asymmetric nuclear matter at the saturation density, $\rho_{0}=0.148$ fm$^{-3}$, are summarized in table~\ref{tab:CCs}.

\section{Numerical results}
\label{sec:results}

Figure~\ref{fig:1} displays the mass{\textendash}radius relations of neutron stars obtained with the OMEG family.
These interactions are constructed so as to satisfy the $2M_{\odot}$ constraint on the maximum neutron-star mass.
Among them, OMEG0 predicts the largest maximum mass, as its EoS is the stiffest at high densities owing to the absence of the quartic $\omega$-meson self-interaction in Eq.~\eqref{eq:NL}.
\begin{figure}
\centering
\includegraphics[width=7.0cm,clip]{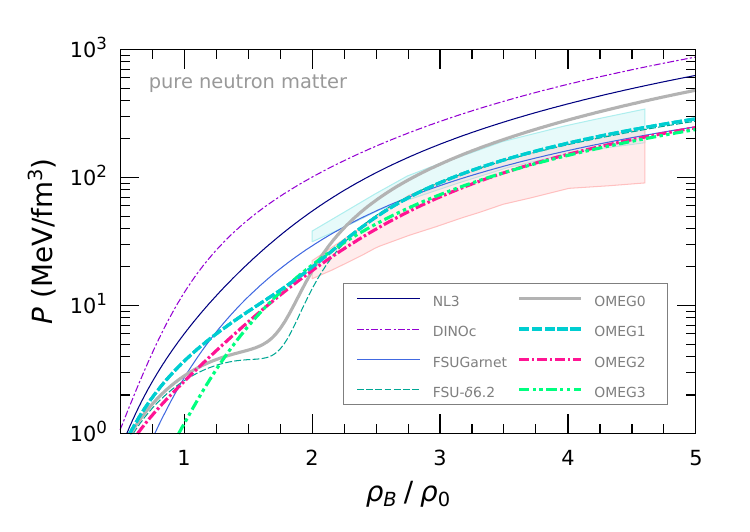}
\caption{Pressure, $P$, in pure neutron matter as a function of baryon density ratio, $\rho_{B}/\rho_{0}$. The empirical constraints on the nuclear EoS extracted from the particle-flow analyses in heavy-ion collisions are also provided~\cite{Danielewicz:2002pu}.}
\label{fig:2}
\end{figure}
\begin{figure*}
\centering
\includegraphics[width=12.0cm,clip]{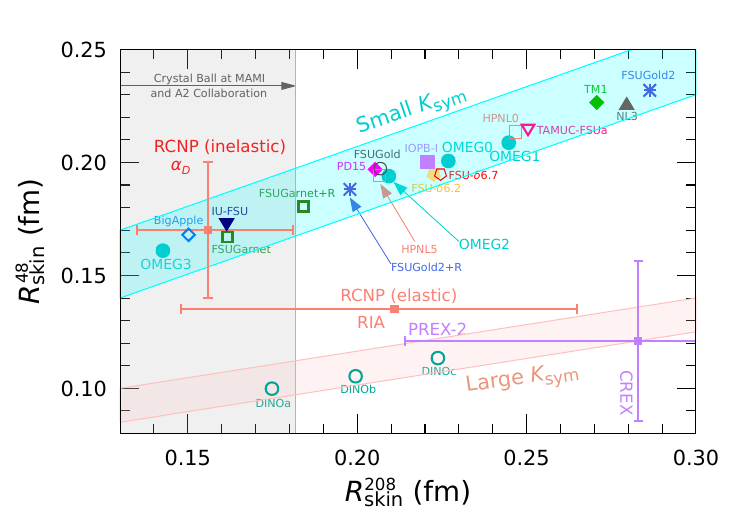}
\caption{Neutron skin thickness of $^{48}$Ca and $^{208}$Pb, $R_{\rm skin}^{48}$ and $R_{\rm skin}^{208}$. A detailed description of all the interactions used in this figure is provided in Ref.~\cite{Miyatsu:2024ioc}.}
\label{fig:3}
\end{figure*}

The OMEG family is specifically designed to yield relatively small neutron-star radii through the inclusion of $\sigma$-$\delta$ mixing.
This mixing softens $E_{\textrm{sym}}(\rho_{B})$ around $\rho_{B}\simeq2\rho_{0}$, leading to more compact stellar configurations~\cite{Miyatsu:2025dxs}.
Consequently, all OMEG parameter sets reproduce the NICER radius measurements remarkably well.
As shown in table~\ref{tab:CCs}, the corresponding dimensionless tidal deformability is also sufficiently small, lying within the GW170817 constraint of $\Lambda_{1.4}=190^{+390}_{-120}$~\citep{LIGOScientific:2018cki}.

Figure~\ref{fig:2} shows the pressure, $P$, of isospin-asymmetric nuclear matter.
In the OMEG family, $P$ remains smaller than that of FSUGarnet---a widely used RMF interaction that reasonably well reproduces the data from terrestrial nuclear experiments and astrophysical observations---up to approximately $2\rho_{0}$.
This reduction, which originates from the softening of $E_{\textrm{sym}}(\rho_{B})$ in this density region corresponding to the core density of a canonical $1.4M_{\odot}$ neutron star, naturally yields smaller $R_{1.4}$ and $\Lambda_{1.4}$, as explained in figure~\ref{fig:1}, thereby allowing the OMEG family to satisfy the NICER and GW170817 constraints.
In practice, the softening of $E_{\textrm{sym}}(\rho_{B})$ and the associated reduction in $P$ are controlled primarily by the curvature parameter, $K_{\textrm{sym}}$, which is negative in all OMEG interactions, as listed in table~\ref{tab:CCs}.

Beyond this intermediate-density region, however, the EoS must stiffen again to support $2M_{\odot}$ neutron stars.
The OMEG interactions exhibit this soft-to-hard evolution: although the EoS remains soft up to around $2\rho_{0}$, $P$ increases rapidly at higher densities while remaining broadly consistent with heavy-ion collision constraints.
Thus, the OMEG family provides a soft EoS at intermediate densities and a sufficiently stiff EoS at high densities within a RMF framework.

In figure~\ref{fig:3}, we summarize the neutron skin thickness of $^{48}$Ca and $^{208}$Pb, $R_{\textrm{skin}}^{48}$ and $R_{\textrm{skin}}^{208}$, using the RMF models, together with various experimental constraints.
The PREX-2 experiment through the parity-violating electron scattering reports a relatively large $R_{\textrm{skin}}^{208}$ ($R_{\textrm{skin}}^{208}=0.283\pm0.071$~fm)~\cite{PREX:2021umo}, whereas the CREX measurement indicates a much smaller $R_{\textrm{skin}}^{48}$ ($R_{\textrm{skin}}^{48}=0.121\pm0.026(\mathrm{exp.})\pm0.024(\mathrm{model})$~fm)~\cite{CREX:2022kgg}.
Additional constraints from the electric dipole polarizability of $^{48}$Ca (RCNP; $R_{\rm skin}^{48}=0.14$--$0.20$~fm)~\citep{Birkhan:2016qkr}, the complete electric dipole response of $^{208}$Pb (RCNP; $R_{\rm skin}^{208}=0.156_{-0.021}^{+0.025}$~fm)~\citep{Tamii:2011pv}, the coherent pion photoproduction cross sections measurement of $^{208}$Pb (MAMI; $R_{\rm skin}^{208}=0.15\pm0.03(\mathrm{stat.})_{-0.03}^{+0.01}(\mathrm{sys.})$~fm)~\citep{Tarbert:2013jze} are also shown in figure~\ref{fig:3}.

A clear correlation is observed between $R_{\rm skin}^{48}$ and $R_{\rm skin}^{208}$: models predicting a larger $R_{\rm skin}^{48}$ also yield a larger $R_{\rm skin}^{208}$, whereas those with smaller $R_{\rm skin}^{48}$ correspondingly predict smaller $R_{\rm skin}^{208}$.
Because of this strong, positive correlation, it is generally difficult for any single parametrization to simultaneously reproduce both the PREX-2 and CREX measurements.

Most conventional RMF interactions are found in the small-$K_{\textrm{sym}}$ region in figure~\ref{fig:3}, typically with $K_{\textrm{sym}}\le100$ MeV, and their predicted $R_{\rm skin}^{48}$ deviate significantly from the very small value measured by the CREX Collaboration.
The OMEG family is constructed specifically to address this situation: OMEG0, OMEG1, and OMEG2 are designed to accommodate the larger $R_{\rm skin}^{208}$, indicated by the PREX-2 experiment, whereas OMEG3 is tuned to reproduce the smaller $R_{\rm skin}^{208}$ extracted from the RCNP electron-scattering data and the smaller $R_{\rm skin}^{48}$ from the CREX measurement, simultaneously.
Although DINOc appears to reproduce both the PREX-2 and CREX results~\cite{Reed:2023cap}, it requires an extremely strong $\delta$-meson coupling, which drives $K_{\textrm{sym}}$ to very large positive values ($\ge500$ MeV) and leads to neutron-star radii far outside the NICER constraints.

\section{Summary}

We have presented a new set of RMF interactions that include the $\sigma$-$\delta$ and $\omega$-$\rho$ mixing, collectively referred to as the OMEG family.
These interactions are constructed so as to satisfy nuclear properties of finite nuclei while also reproducing the astrophysical constraints on neutron-star radii and tidal deformabilities obtained from NICER and GW170817.
The $\sigma$-$\delta$ mixing softens $E_{\textrm{sym}}(\rho_{B})$ around $\rho_{B}\simeq2\rho_{0}$, which in turn reduces $P$ of isospin-asymmetric nuclear matter and naturally yields small values of $R_{1.4}$ and $\Lambda_{1.4}$.
At higher densities, however, the EoS becomes sufficiently stiff to support neutron stars with masses $\gtrsim$ $2M_{\odot}$.

Although it remains challenging for any single parametrization to simultaneously reproduce the PREX-2 and CREX measurements, the OMEG family can accommodate the large $R_{\textrm{skin}}^{208}$ suggested by the PREX-2 Collaboration while still explaining the small neutron-star radii and tidal deformabilities inferred from astrophysical observations.
This behavior is intimately connected to $K_{\textrm{sym}}$, which takes negative values in all OMEG interactions and drives the characteristic soft-to-hard evolution of the nuclear EoS.

\section*{Acknowledgements}

This work was supported by the Basic Science Research Program through the National Research Foundation of Korea (NRF) under Grant Nos. RS-2025-16066382, RS-2025-16071941, RS-2023-00242196, and RS-2021-NR060129.

%
%
%

\end{document}